# Equivalent-Time-Active-Cavitation-Imaging Enables Vascular-Resolution Blood-Brain-Barrier-Opening-Therapy Planning

**Authors** – Samuel Desmarais[1], Gerardo Ramos-Palacios[2], Jonathan Porée[1], Stephen A. Lee[1], Alexis Leconte[1], Abbas F. Sadikot[2], Jean Provost[1,3]; [1]Polytechnique Montréal, Montréal, Canada; [2]Montreal Neurological Institute and Hospital, McGill University, Montréal, Canada; [3]Institut de Cardiologie de Montréal, Montréal, Canada

Samuel Desmarais: samuel.desmarais@polymtl.ca, https://orcid.org/0009-0000-5886-7748

Gerardo Ramos-Palacios: gerardo.ramos@mail.mcgill.ca, https://orcid.org/0000-0003-0861-6152

Jonathan Porée: jonathan.poree@polymtl.ca, https://orcid.org/0000-0002-4179-9264

Stephen A. Lee: stephen.lee@polymtl.ca, https://orcid.org/0000-0002-1315-4266

Alexis Leconte: alexis.leconte@polymtl.ca, https://orcid.org/0009-0001-2420-9250

Abbas F. Sadikot: abbas.sadikot@mcgill.ca, https://orcid.org/0000-0003-1491-2038

Jean Provost: **Corresponding Author**, jean.provost@polymtl.ca, https://orcid.org/0000-0003-2057-2199

## ABSTRACT

Linking cavitation and anatomy was found to be important for predictable outcomes in Focused-Ultrasound Blood-Brain-Barrier-Opening and requires high resolution cavitation mapping. However, cavitation mapping techniques for planning and monitoring of therapeutic procedures either 1) do not leverage the full resolution capabilities of ultrasound imaging or 2) place strong constraints on the length of the therapeutic pulse. This study aimed to develop a high-resolution technique that could resolve vascular anatomy in the cavitation map.

Herein, we develop BP-ETACI, derived from bandpass sampling and dual-frequency contrast imaging at 12.5 MHz to produce cavitation maps prior and during blood-brain barrier opening with long therapeutic bursts using a 1.5-MHz focused transducer in the brain of C57BL/6 mice.

The BP-ETACI cavitation maps were found to correlate with the vascular anatomy in ultrasound localization microscopy vascular maps and in histological sections. Cavitation maps produced from non-blood-brain-barrier disrupting doses showed the same cavitation-bearing vasculature as maps produced over entire blood-brain-barrier opening procedures, allowing use for 1) monitoring FUS-BBBO, but also for 2) therapy planning and target verification.

BP-ETACI is versatile, created high resolution cavitation maps in the mouse brain and is easily translatable to existing FUS-BBBO experiments. As such, it provides a means to further study cavitation phenomena in FUS-BBBO.

## INTRODUCTION

FOCUSED-Ultrasound Blood-Brain-Barrier-Opening (FUS-BBBO) with microbubbles (μB) was first observed in rabbits [1] and is an emerging therapeutic tool under clinical trial that has the potential to lessen the impact of neurological diseases such as Alzheimer's disease, Parkinson's disease, and brain cancer [2]. The technique is promising because it allows known large molecular weight therapeutic agents to cross the BBB and be effective in the brain.

Other techniques have been studied for promoting the action of diverse pharmacological agents in the central nervous system across the BBB. These techniques include intrathecal injections [3], [4] pharmacological agents that promote BBB permeability [5], and delivery of vectors that readily cross the BBB [6], [7]. FUS-BBBO is of particular interest because focused ultrasound allows the technique to be less invasive and its effect limited to the region of highest intensity in the emitted pressure field.

The BBB disrupting effect of FUS comes from the action of ultrasound on contrast agents composed of μB administered intravenously. As μB circulate and are exposed to the rapidly oscillating pressure in the acoustic field, they undergo rapid changes in size without bursting, a phenomenon termed non-inertial cavitation, which induces intense mechanical and fluidic processes in the immediate surroundings and triggers transport across the BBB [8].

The localized effect of FUS-BBBO requires targeting (1) to ensure that the brain regions requiring treatment receive the therapeutic agent and (2) to minimize off-target delivery. This is especially important because interaction between different tissue structures and FUS leads to different outcomes depending on the target [9], [10]. Current methods of visualizing the opening include post-therapeutic magnetic resonance imaging (MRI) and concurrent acoustic mapping. MRI of gadolinium in the brain parenchyma is used [1] to verify the extent of BBBO because the gadolinium compounds used as contrast agents in MRI do not cross the undisturbed BBB. Acoustic monitoring is used in FUS-BBBO to minimize damage because inertial and non-inertial cavitation have particular backscatter signatures which are used to distinguish and quantify them [11]. Acoustic mapping approaches use a multi-element imaging transducer to receive scattered signals during FUS sonication and beamform maps of the acoustic activity. Passive Acoustic Mapping (PAM) is one such approach but its axial resolution is limited by the use of continuous mode therapeutic sonication [12]. Imaging transducers can also be used as the receiving component in a pulse-echo ultrasound system where the pulse is provided by the focused transducer. Paired with the use of few-cycles therapeutic pulses comparable to those used in ultrasound imaging, the technique produced B-mode images of the cavitation activity [13]. Therapy and cavitation mapping was also conducted together from the imaging probe [14]. However, to our knowledge, no cavitation mapping technique allows for in vivo vascular-resolution cavitation imaging without requiring very short therapeutic pulses.

Recently, it has been shown that it is possible to map non-inertial cavitation at high resolution using Equivalent-Time Active-Cavitation Imaging (ETACI), Phantom studies demonstrated the correlation between the ETACI signal and pressure levels and showed initial results in vivo [15]. ETACI is a method based on Radial Modulation (RM) where Doppler spectral replicates are generated in the pixels undergoing cavitation. Briefly, RM methods typically use a two-frequency approach where a low-frequency wave is used to modulate the size of µB and thus their scattering characteristics, and a high-frequency pulse is used to image the µB at different phases of the RM cycle.

RM has been used mostly as a µB-contrast imaging technique because the scattering-modulation effect is specific to µB. Work on RM first showed the change in amplitude of a high-frequency wave scattered by microbubbles through RM cycles induced by a low-frequency wave [16] which has been used to measure the size of individual µB [17]. µB were found to produce different backscattered signal when high-frequency imaging pulses imaged them in compression and expansion phases of the RM [18]. Second order ultrasound field (SURF) imaging used dual-band pulses emitted by the imaging probe to image µB in compression and expansion in vivo to conduct contrast imaging [19].

Recently, ultrafast imaging was used to develop contrast RM techniques in phantoms. Ultrafast radial modulation imaging (uRMI) [20] increased the number of sample phases at which the RM cycle is sampled by leveraging ultrafast image formation used demodulation to produce contrast images. Further work identified that RM could be induced by low frequency ultrasound (100 kHz) at depth and that Singular Value Decomposition (SVD)-based extraction of the RM signal was possible in phantoms, making RM more robust to tissue motion [21]. [20], [21] and [15] found artifacts in the form of additional peaks in the Doppler spectrum, which remained unexplained.

The objective of this study was to establish the potential of ETACI to reliably map the non-inertial cavitation produced by a therapeutic transducer at high resolution in the mouse brain in vivo and support BBBO therapy planning. We developed and validated a novel framework for ETACI based on bandpass sampling (BP-ETACI), that solved the spectral artifacts, enabled further uncoupling of the framerate and the modulation frequency, simplifying the method, increasing adaptability by allowing the selection of the aliasing frequency in the received signal, and improving image quality by eliminating spectral artifacts. In vitro, we validated the selectivity of BP-ETACI towards cavitating µB and explored how modulation and Doppler mix. We then illustrated the potential of BP-ETACI by demonstrating that cavitation can be mapped in vivo in mouse models transcranially in the context of FUS-BBBO.

## METHODS

### Theoretical Framework

First, we develop the equations that describe the framework. Then, we use these equations to trace the shape of the spectrum that would be obtained with the imaging and modulation parameters selected for this study.

#### *Pulsed-Doppler*

For pulsed-Doppler (PD), the signal of interest comes from the movement of scatterers between pulse-echo events. As such, the received signal in slow time can be modeled with (1) for a scatterer moving towards the probe at speed $v_z$, in a medium where the speed of sound is $c$, and observed with $M$ cycle pulses of center frequency $f_0$.

$$d(t) = h(t) \exp(-i2\pi f_d t) \quad (1)$$

Where $h(t)$ is a window of length $|Mc/f_0 v_z|$, which corresponds to the transit time of the scatterer as observed by the system at a fixed depth. The motion of the scatterer is observed as a frequency $f_d = 2v_z f_0/c$ in slow time. This frequency shift is shown in Fig. 1. A where the spread of the frequency band is attributed to scatterer motion. [22]

### BP-ETACI Modulation

In BP-ETACI, the scatterers are μB and their backscatter amplitude is modulated externally. The received signal can thus be modeled with (2) in which $A$ and $B$ parametrize this modulation.

$$s(t) = d(t)[A + B \cos(2\pi f_m t)] \quad (2)$$

Combining, (1) and (2), we obtain

$$s(t) = h(t) \exp(-i2\pi f_d t) [A + B \cos(2\pi f_m t)] \quad (3)$$

The received signal $s(t)$ develops to a sum of three scaled and gated complex exponentials.

$$s(t) = A\, h(t) \exp(-i2\pi f_d t) + \frac{B}{2} h(t) \exp(-i2\pi (f_m + f_d)t) + \frac{B}{2} h(t) \exp(-i2\pi (-f_m + f_d)t) \quad (4)$$

In the frequency domain, these complex exponentials yield three signal bands around $f = 0$, $f = f_m$ and $f = -f_m$.

$$S(f) = A\pi\, H(f) * \delta(f - f_d) + \frac{B\pi}{2} H(f) * \delta\big((f - f_m) - f_d\big) + \frac{B\pi}{2} H(f) * \delta\big((f + f_m) - f_d\big) \quad (5)$$

The modulating pressure field is not spatially uniform, as it originates from a focused transducer in FUS-BBBO. We can rewrite the received signal as

$$S(f) = A\pi\, H(f) * \delta(f - f_d) + \frac{\pi}{2} B(r)\, H(f) * \delta\big((f - f_m) - f_d\big) + \frac{\pi}{2} B(r)\, H(f) * \delta\big((f + f_m) - f_d\big) \quad (6)$$

With $B(r)$ the spatial distribution of the focused transducer's pressure field. Thus, the last two bands of $S(f)$ contain information about the spatial distribution of the focused transducer's pressure field. Fig. 1. B shows the two bands added by modulating the scatterers amplitude over time. The central band (orange) does not contain $B(r)$ in its expression whereas the two side-bands (blue) do contain $B(r)$ in their expression. Thus, the central band, which contains unmodulated PD signal from all moving structures, not only from modulated μB, is spurious when mapping cavitation.

### Sampling

Using bandpass sampling, one can extract $B(r)$ even when using a framerate $f_s$ much smaller than the modulation frequency. Indeed, for a signal sampled at a sampling frequency $f_s$, the baseband spectrum contained between $-f_s/2 \leq f < f_s/2$, is the superposition of all bands $-f_s/2 \leq f - k f_s < f_s/2$ for all $k \in \mathbb{Z}$. Thus, the observed spectrum $S_o(f)$ is

$$S_o(f) = \sum_{k=-\infty}^{\infty} S(f) * \delta(f - k f_s) \quad (7)$$

Applying a change of variables to obtain the spectrum between $-1/2 \leq f/f_s < 1/2$ gives

$$S_o\left(\frac{f}{f_s}\right) = \sum_{k=-\infty}^{\infty} S\left(\frac{f}{f_s}\right) * \delta\left(\frac{f}{f_s} - k\right) \quad (8)$$

As such, the received signal contains the three signal bands of (6) between $-1/2 \leq f/f_s < 1/2$

$$S_o\left(\frac{f}{f_s}\right) = S\left(\frac{f}{f_s}\right) * \delta\left(\frac{f}{f_s} - k_1\right) + S\left(\frac{f}{f_s}\right) * \delta\left(\frac{f}{f_s} - k_2\right) + S\left(\frac{f}{f_s}\right) * \delta\left(\frac{f}{f_s} - k_3\right) \quad (9)$$

With $k_1 = 0$, $k_2$ and $k_3$ obeying these constraints:

$$\{1/2 \leq 0/f_s - k_1 < 1/2\}\,; \{1/2 \leq f_m/f_s - k_2 < 1/2\}\,; \{1/2 \leq -f_m/f_s - k_3 < 1/2\} \quad (10)$$

Thus, choosing $f_m$ and $f_s$ so their ratio is not whole separates the signal bands that contain $B(r)$ from the central band, which is fixed at 0 Hz. Fig. 1. illustrates the possibility of generating separated bands (C) or superimposed bands at 0 Hz (D).

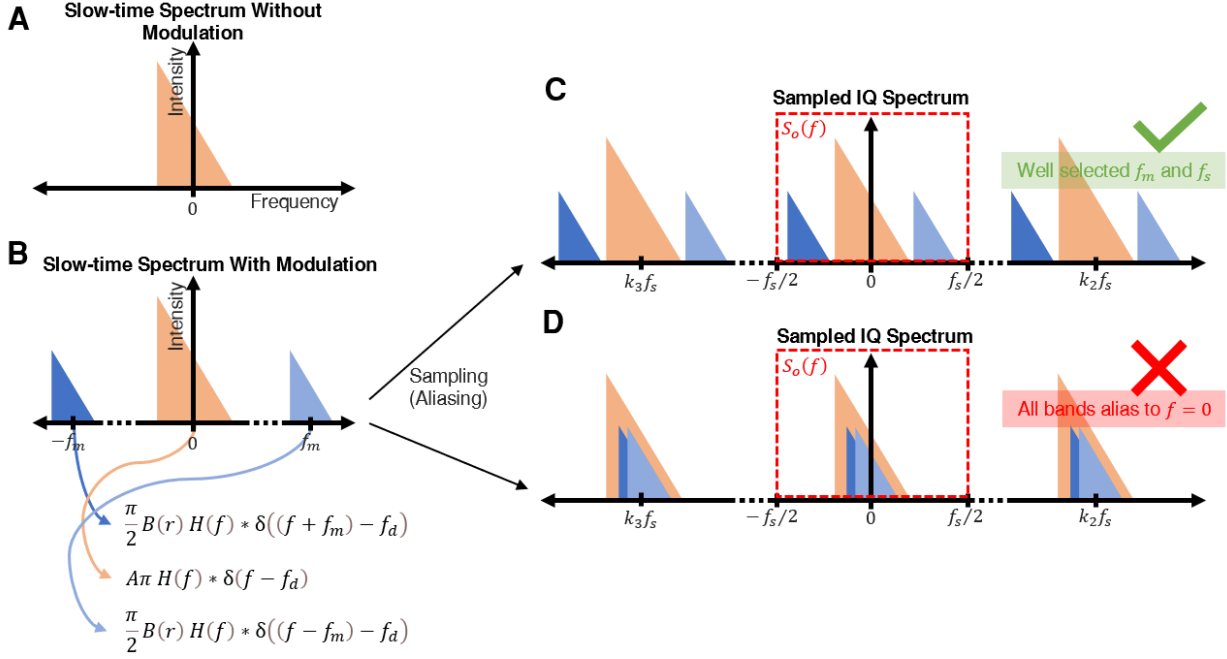

Fig. 1. Spectral View of BP-ETACI Signal Formation. (A) Slow-time-wise spectrum of moving µB. (B) Slow-time spectrum containing the pulsed-Doppler band at $f = 0$ (orange) and modulated bands at $f = -f_m; f_m$ (blue). (C) Slow-time spectrum that contains non-overlapping bands between $-f_s/2$ and $f_s/2$. (D) Slow-time spectrum with overlapping bands between $-f_s/2$ and $f_s/2$.

### BP-ETACI Demodulation

Information on $B(r)$ is only present in the spectral bands created by the modulation. Thus, to isolate this information, the slow time-sampled data $s(t)$ must be demodulated at the modulation frequency. The width of the demodulation low-pass (LP) filter is chosen so that only one band of the received signal is preserved. Demodulating around the $f = f_m$ band gives

$$g(t) = \text{LP}\left[s(t) \exp\left(i2\pi \left(\frac{f_m}{f_s} - k_2\right)t\right)\right]$$

$$= \tfrac{1}{2}B(r)h(t)\exp\left(-i2\pi\left(\frac{f_d}{f_s}\right)t\right) \quad (11)$$

When sufficient data is acquired, averaging temporally reduces (11) to

$$\hat{B}(r) = q(r)B(r) \quad (12)$$

Where $q(r)$ is the spatial distribution of µB. $\hat{B}(r)$ is thus a quantity that reflects the number of microbubbles and the amplitude of their non-inertial cavitation, which is directly related to the local pressure.

### Validation of the Framework

The spectrum obtained from (4) was traced to validate the separation between modulated and unmodulated bands and to understand how both the motion signal (PD) and the modulation signal interacted to produce the resulting signal.

Specifically, equation (4) was sampled at a single location and a rate of $f_s = 4$ kHz with the following parameters: $c = 1540$ m/s, $M = 3$, $f_0 = 12.5$ MHz, $f_m = 1.50132$ MHz, $A = 1$ and $B = 0.1$, $v_z = 5; 10; 15$ mm/s and $h(t)$ a Hann window. Doing so, a 1D temporally sampled signal vector was obtained. The spectrum of this signal was then computed. In this situation, $\pm f_m/f_s = \pm 375.33$, $k_2 = 375$ and $k_3 = -375$ thus we expect the modulated bands to be centered at $f/f_s = \pm 0.33$ to satisfy (10). To model multiple scatterers, the process was repeated and averaged while maintaining all the parameters constant, except for the velocity, which was uniformly selected along the radius according to Poiseuille flow (radius = 2.5mm, $v_{z\,max} = 5; 10; 15$ mm/s).

## Experimental Setup

For imaging, a programmable ultrasound scanner (Vantage 256, Verasonics) was used to control a linear imaging probe (L8-18i, GE). For RM and BBBO therapy, a focused monoelement transducer (H-234, Sonic Concepts) was driven by an RF amplifier (Empower RF) with a signal generated by a function generator (SDG6000, Siglent).

The focused transducer was mounted transducing-face down in a 3D-printed conical water reservoir. A silicone cast of the imaging transducer was used to mount it in a way that placed the focal zone of the focused transducer in the imaging plane of the probe. Both the focused transducer and the imaging probe were mounted to a 3D-printed bracket. The reservoir was filled with room-temperature water and a latex membrane was used to seal the bottom. Fig. 2 shows a cross-section view of the assembly.

## Imaging

The imaging sequences used for this study are detailed here. The ultrasound scanner emitted 3-cycle pulses centered at 12.5 MHz in a plane wave imaging scheme. Transmission and reception occurred only on the 128 middle elements of the imaging probe. For all acquisitions, the reception bandwidth was 100%, frames were acquired in ensembles of 1200 with pauses between each ensemble. Frames were DAS beamformed and plane wave angles were coherently compounded.

For ULM acquisitions, the frame rate was 1.6 kHz with 5 angles per frame. The focused transducer was kept off. An SVD-based clutter filter was applied to each ensemble to highlight circulating μB [23] and a spatiotemporal localization algorithm [24] was used to extract super-resolved μB trajectories. These trajectories were accumulated on a high-resolution grid to obtain maps of μB density.

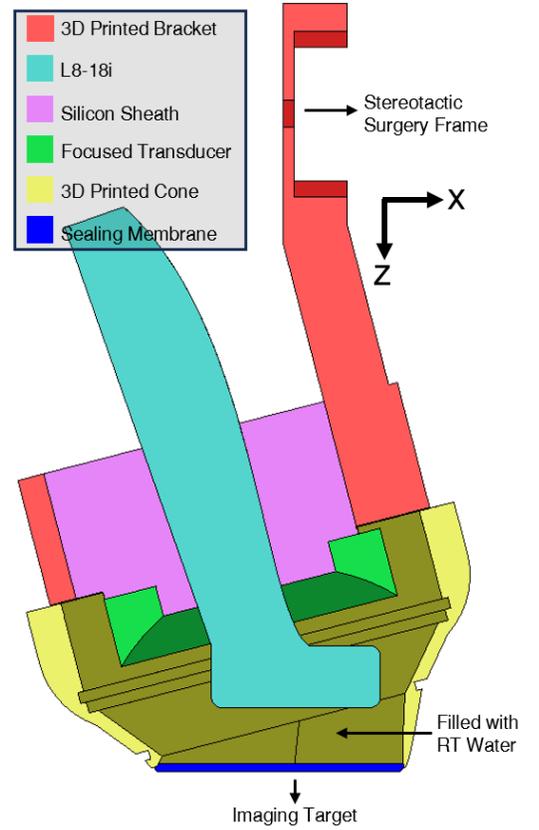

Fig. 2. Cut view of the experimental assembly used for all acquisitions.

For BP-ETACI acquisitions, the frame rate was $f_s = 4$ kHz with 4 planewave angles per frame. The focused transducer emitted 300 ms tone bursts at $f_m = 1.50132$ MHz with a PRF of 1 Hz. Here, $f_m/f_s = 375.33$ thus modulated bands are expected at $f/f_s = \pm 0.33$ so, after coherent compounding, each ensemble was demodulated in slow time according to (11) at a normalized frequency of $f/f_s = 0.33$. The demodulation filter used in (11) was a frequency-domain Blackman-Harris window of width 1.6 kHz. The recording of each ensemble was triggered on the start of a tone burst. Some strong clutter remained at the skull and was removed by removing the first two singular values of the Casorati matrix formed with the middle 1000 frames of the ensemble [23]. Then, the magnitude of each pixel was averaged temporally to get a spatial map of the cavitation induced by the focused transducer.

## In Vitro Experiments

### *Free-Field Experiment*

BP-ETACI was used to map the focal zone of the focused transducer. The experimental assembly was mounted above a tank filled with a 14 μL/L dilution of contrast agent (Definity, Lantheus) in room temperature water agitated with a magnetic stirrer (Fig. 4. A). 60s of BP-ETACI data were acquired. The total acquisition time was chosen to ensure μB visited the field of view uniformly, under which conditions the resulting map is proportional to the spatial distribution of the pressure induced by the focused transducer. The focused transducer was driven at very low amplitude (0.01 MPa) for this experiment to limit the radiation force and avoid pushing the μB, as significant PD frequencies can be achieved from this push and could be confused with modulation.

### *Peak Negative Pressure Field Map*

The focal zone was mapped with a hydrophone to verify the BP-ETACI cavitation map. The experimental assembly was mounted on a bracket and submerged in a tank filled with room-temperature water. A hydrophone (Y-104, Sonic Concepts) was mounted on a 3-axis robot (X-LSM200A-E03, Zaber) and connected to a data acquisition device (5242D, Picoscope). The minimum value of each pressure waveform was extracted and converted to a peak negative pressure map.

*Channel Phantom Experiment*

We verified that BP-ETACI only detects cavitating µB, not tissue and µB outside the focal zone. The experimental assembly was mounted and aligned above a channel phantom, with the probe imaging along the length of the channel (Fig. 5. A). The channel was angled at 8° from the horizontal to allow for an axial component in µB speed. A solution of 56 µL/L Definity contrast agent was prepared by dilution with water. The channel diameter was 2.5 mm. The focused transducer was driven at 0.02 MPa. The flow rate was set to 5 mL/min and 6s of BP-ETACI data were acquired. The slow-time spectrum of each pixel was computed and then averaged over three subregions of the field of view: 1) inside the channel and the focal spot, 2) inside the channel, but outside the focal spot, and 3) inside the focal spot but outside the channel in the tissue portion of the phantom.

With the same setup, we verified that BP-ETACI produces similar cavitation maps under various flow rates. BP-ETACI cavitation maps were acquired at flow rates of 5, 10, 20 mL/min, and at a fourth, very fast, flow rate achieved by pushing the µB solution manually. This fourth flow rate was selected because it made the central PD band overlap the filter passband in (11). For each experimental condition, 6 s of BP-ETACI data were acquired. Before demodulating according to (11), the spectrum of each pixel was computed and averaged over a group of pixels spanning the width of the channel inside the focal spot for each flow rate.

**In Vivo Experiments**

We used BP-ETACI to map the cavitation transcranially in the mouse brain prior to and during FUS-BBBO treatments. The study group was composed of 6 wild type C57BL/6 female mice and was split into two groups of 3 animals. BP-ETACI cavitation maps were produced at low cavitation doses for each animal in the first group, termed the mapping-only-group. The second group, termed the BBBO-group, underwent a BBBO therapy procedure and BP-ETACI data was acquired during the whole procedure. This served two purposes: 1) establishing that BP-ETACI can map the cavitation transcranially in the mouse brain at doses that do not open the BBB, and 2) establishing that BP-ETACI reliably maps regions where the BBB is opened by FUS-BBBO therapy.

The mice were deeply anesthetized using a mix of 1L/min of oxygen and isoflurane 5% for induction and maintained at 1.5-2.5% for the duration of the procedure. Their heads were immobilized in a stereotactic frame (Model 940, Kopf), were shaved, and were depilated. Their temperature was maintained with a warm-water-circulation blanket. Sterile saline solution 0.9 % was administered intra-peritoneally at 0.5 mL per 10g of body weight to maintain adequate intravascular volume during the procedure. Room-temperature deionized water was applied to the skin and to the outside of the experimental assembly's water retaining membrane. Ultrasound coupling gel was then applied to the mouse head, the experimental assembly was mounted to the stereotactic system, and lowered above the head to provide coronal imaging planes (Fig. 9. A). The acoustic coupling between the assembly and the animal was verified with B-mode imaging.

All mice underwent an ULM imaging protocol prior to BP-ETACI cavitation mapping. For ULM imaging, 5 minutes of data was recorded over 15 minutes. A bolus consisting of a 4 µL/g of bodyweight dose of 1:20 phosphate buffered saline (PBS)-diluted Definity solution was injected in the tail vein of the animal following the start of the recording.

BP-ETACI was conducted after the previously injected µB were allowed to decay. For all in vivo BP-ETACI experiments, one bolus consisting of a 4 µL/g dose of 1:20 PBS-diluted Definity solution and of a 2 µL/g dose of 1% Evans blue dye in PBS was used. The focused transducer was set to a peak-negative-pressure of 0.1 MPa. For animals in the mapping-only-group, the focused transducer was set to emit 5 bursts and thus 5 ensembles were recorded. The bolus was injected in the tail vein of the animal 10 seconds before the start of the sonication. For the animals in the BBBO-group, real time BP-ETACI cavitation mapping and focused transducer sonication were initiated, and bolus injection occurred within 30 seconds from the start of the sonication. The arrival of the bolus was observed in BP-ETACI. The real-time BP-ETACI processing was then done for 1 therapeutic burst every 20 to monitor that the focal zone placement was maintained. Recording and focused transducer sonication were continued for a total of 4 minutes after bolus injection. The total time of BP-ETACI data recorded during BBBO varied between animals depending on the save and process time of ensembles. Some of the focused transducer bursts that occurred late into each BBBO procedure were not recorded because their trigger signal overlapped with the save time of the previous burst. From all the ensembles acquired during BBBO, two cavitation maps were produced. One from the first 5 focused transducer bursts acquired later than 10s after the bolus injection and one from all ensembles acquired later than 10s after the bolus injection.

The mice were transcardially perfused with 4% paraformaldehyde at 7.4 pH 24h after the procedure. The brains were cryoprotected, frozen, and cut into 40 µm coronal sections. The section that best aligned with the focal zone for each animal was identified and tile scanned with a fluorescence microscope (Aperio Versa 200, Leica). The same exposure settings were used for each slide.

## RESULTS

### Illustration of the Theoretical Framework

Simulations of the signal framework were conducted to evaluate the capability of BP-ETACI to separate the BP-ETACI signal, originating from the modulation induced by the focused transducer, from the PD signal, originating from scatterer motion. This separation was evaluated in the frequency domain. The signal spectrum was simulated for single scatterers and Poiseuille flow. We show that BP-ETACI can separate modulation from flow in the frequency domain. For the single scatterer model, the frame rate and the modulation frequency used in the simulation were chosen because at those frequencies the signal framework in (6) places the center of the modulated spectral bands at $0 \neq f/f_s = 0.33, -0.33$, away from the PD signal at $f/f_s = 0$. Fig. 3. A shows simulated spectra for single scatterers moving at different speeds. The curves show modulated sidebands around $f/f_s = 0.33$ and $-0.33$, as predicted. Scatterer movement shifts all signal bands off their $v = 0$ m/s frequency, shown with dotted lines. Increasingly fast movement increases this shift and broadens the bands, which was predicted by (6). Fig. 3. B shows simulated spectra for an ensemble of 1000 particles in a simulation of Poiseuille flow.

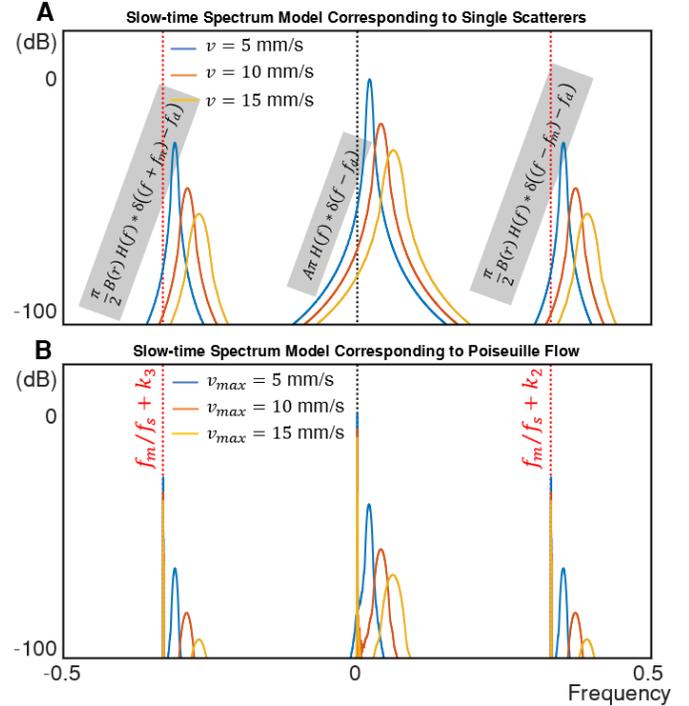

Fig. 3. Slow-time spectra predicted by the developed signal framework. (A) Slow-time spectrum of scatterers moving at different speeds. The parameters used are realistically attainable with ultrasound equipment. (B) Slow-time Spectrum averaged over 1000 scatterers with speeds distributed according to a Poiseuille Flow. All other parameters are the same as in (A).

The signal is concentrated in the same three spectral bands as in Fig. 3. A. A Poiseuille flow produced spectral bands that hold two peaks: one at the $v = 0$ m/s frequency corresponding to scatterers in no-slip-condition along the walls, and one corresponding to flowing scatterers, increasingly offset from the first one as the flow rate increases.

### In Vitro Study

#### *Shape of the Focal Zone*

Prior to demodulation according to (11), the B-mode frames showed μB in the field of view (Fig. 4. B). After demodulation, only the μB inside the focal zone were distinguishable (Fig. 4. C). The temporal accumulation of 60s of demodulated BP-ETACI frames showed the shape of the focal zone (Fig. 4. D) and presented strong similarities with a map of the focused transducer's peak negative pressure as measured with a hydrophone (Fig. 4. E) that was registered to the same field of view as Fig. 4. D. Both the BP-ETACI map and the hydrophone map showed a clearly defined focal zone in the same location and with similar sizes.

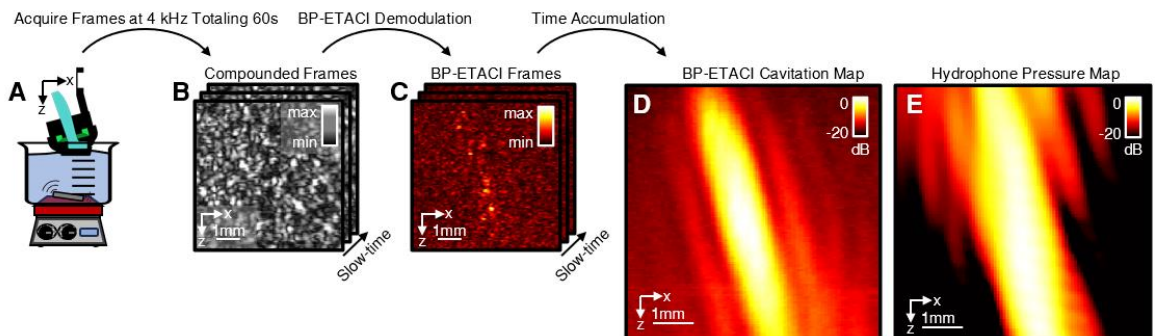

Fig. 4. Mapping the acoustic amplitude of the focused transducer's emission in free field. (A) Plane wave acquisition of agitated dilute microbubbles. (B) Beamforming and coherent compounding. (C) Demodulation and filtering as described in (11) highlights microbubbles inside the focal. (D) Cavitation map obtained by accumulating BP-ETACI images over a long timeframe to ensure the agitated microbubbles visited every point of the field of view uniformly. (E) Peak Negative Pressure map over the imaging plane obtained by raster-scanning the focal spot with a hydrophone. Both the raster scan and the BP-ETACI-obtained map give similar results.

*Selectivity of BP-ETACI Towards Cavitating µB*

While the B-mode frames showed the whole channel phantom (Fig. 5. B), the BP-ETACI cavitation map only held signal in the part of the µB-containing-phantom's channel intersecting the focal zone (Fig. 5. C). Evaluating the average spectrum in sections of the field of view (Fig. 5. D-E-F) revealed that modulated bands appeared strongest at the overlap of the focused transducer's focal zone and of the phantom's channel. Both the signal spectrum in the overlap between the phantom's tissue mimicking material and the focused transducer's focal and the signal spectrum in the phantom's channel outside the focused transducer's focal contained almost no modulated bands.

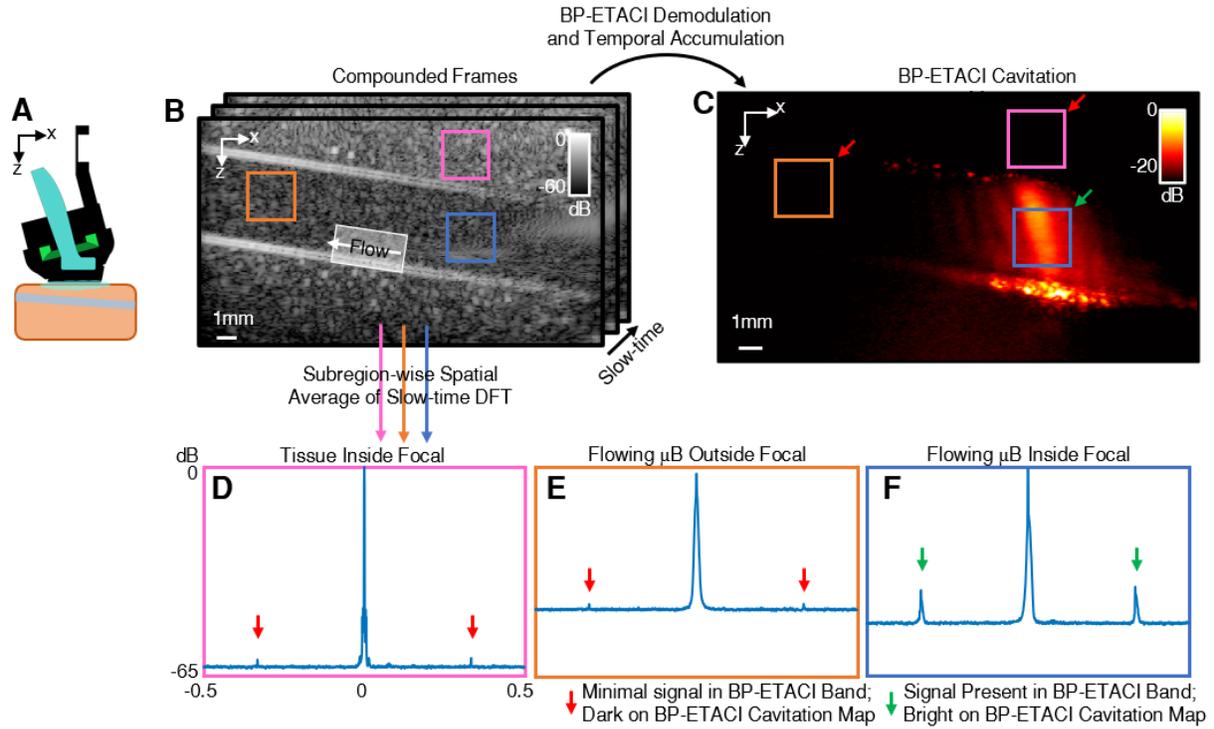

Fig. 5. Cavitation mapping with BP-ETACI when part of the focal spot is obstructed by tissue. (A) Ultrafast plane wave acquisition of dilute microbubbles flowing in a flow phantom. (B) beamforming and coherent compounding. (C) BP-ETACI cavitation map showing the focal spot interrupted by the phantom walls (D-E-F) Double-sided-slow-time-DFTs averaged over subregions of the field of view.

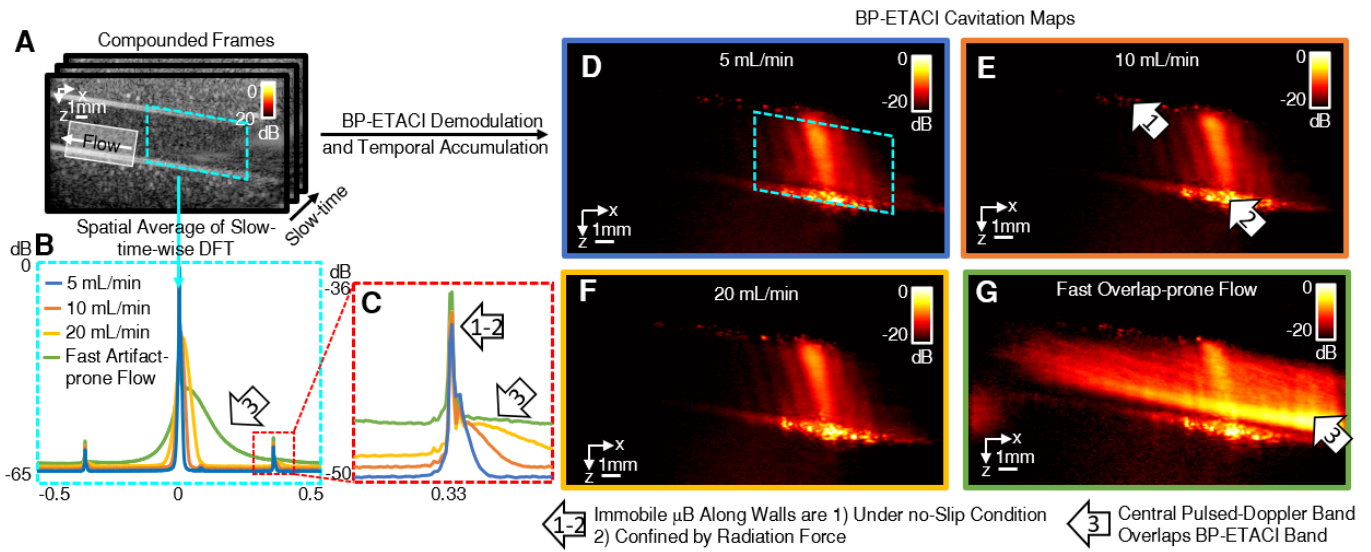

Fig. 6. Cavitation mapping with ETACI at increasingly fast flow rates. (A) Compounded frames of dilute microbubbles flowing in a flow phantom. (B) Double-sided-slow-time-DFTs of increasingly fast flow. (C) Zoom over the BP-ETACI demodulation signal band. (D-E-F) Similar cavitation maps obtained at increasingly fast flow rates. (G) Cavitation map when the flow is fast enough to mix the signal bands.

### *Effect of Flow on BP-ETACI*

In Fig. 6., the influence of flow on BP-ETACI cavitation maps was evaluated in a flow phantom. The spectrum evaluated on B-mode frames (Fig. 6. A) showed BP-ETACI bands for all flow rates (Fig. 6. B-C). The BP-ETACI bands were composed of 1) A tall narrow peak at $f/f_s = 0.33$ which corresponded to μB confined to the phantom's walls. 2) As the flow rate was increased the signal bands corresponding to moving μB shifted in frequency and widened. After demodulation according to (11) and time accumulation, cavitation maps contained the signature of both immobile and flowing μB (Fig. 6. D-E-F-G). The immobile μB were confined by the slow flow near the walls and most likely by radiation forces on the wall away from the focused transducer. Cavitation maps produced with significantly different flow rates were virtually identical (Fig. 6. D-E-F). When a threshold flow rate was reached, the central PD band was wide enough to be included in the filter passband in (11) and was visible in the cavitation map (Fig. 6. G). Bright streaks associated with fast flow appeared over the whole channel.

## In Vivo Study

### *BP-ETACI Produces Cavitation Maps In Vivo*

In vivo, focused-transducer-induced cavitation added a shape that matched the extent of the expected focal zone to BP-ETACI-produced images (Fig. 7. A C) and produced modulated bands in the spectrum (Fig. 7. B D). Additionally, there were noticeable details in the cavitation map along the focal zone, with bright blood vessels and dark low-vascular density regions like the corpus callosum and the ventricular system. The BP-ETACI signal was strongest in the cortex and in subcortical regions. Signal associated with high blood velocities that overlap the passband of the demodulation filter in (11) (Fig. 7. B D) were present bilaterally, whether the focused transducer was turned ON or OFF.

### *BP-ETACI Cavitation Maps can be Obtained Without BBBO*

For the animals in the in the BBBO-group, all the anatomical structures that were visible in the cavitation maps created from data acquired over the full BBBO were also visible in the cavitation maps created from only the first 5 focused transducer bursts that occurred 10 s after μB injection, a non-BBB opening dose, even if noise was reduced when more data was used (Fig. 8. A-F). The cavitation map pixel values (Fig. 8. G-I) obtained at the no-BBBO dose tended to be higher than the pixel values obtained over the BBBO procedure, which was expected for noisy pixels, but this trend was even stronger for pixels in the vasculature. The pixels that deviated the most from the $y = x$ line, like those inside the cyan ellipse in Fig. 8. I, corresponded to cavitation bearing pixels in the cortical and subcortical regions of the corresponding cavitation map. The intensity of the cavitation signal was reduced over time as the µB were eliminated by the mice's metabolism (Fig. 8. J-L). Thus, 5 focused transducers bursts were sufficient to produce cavitation maps and integrating over the whole BBBO-procedure reduced noise.

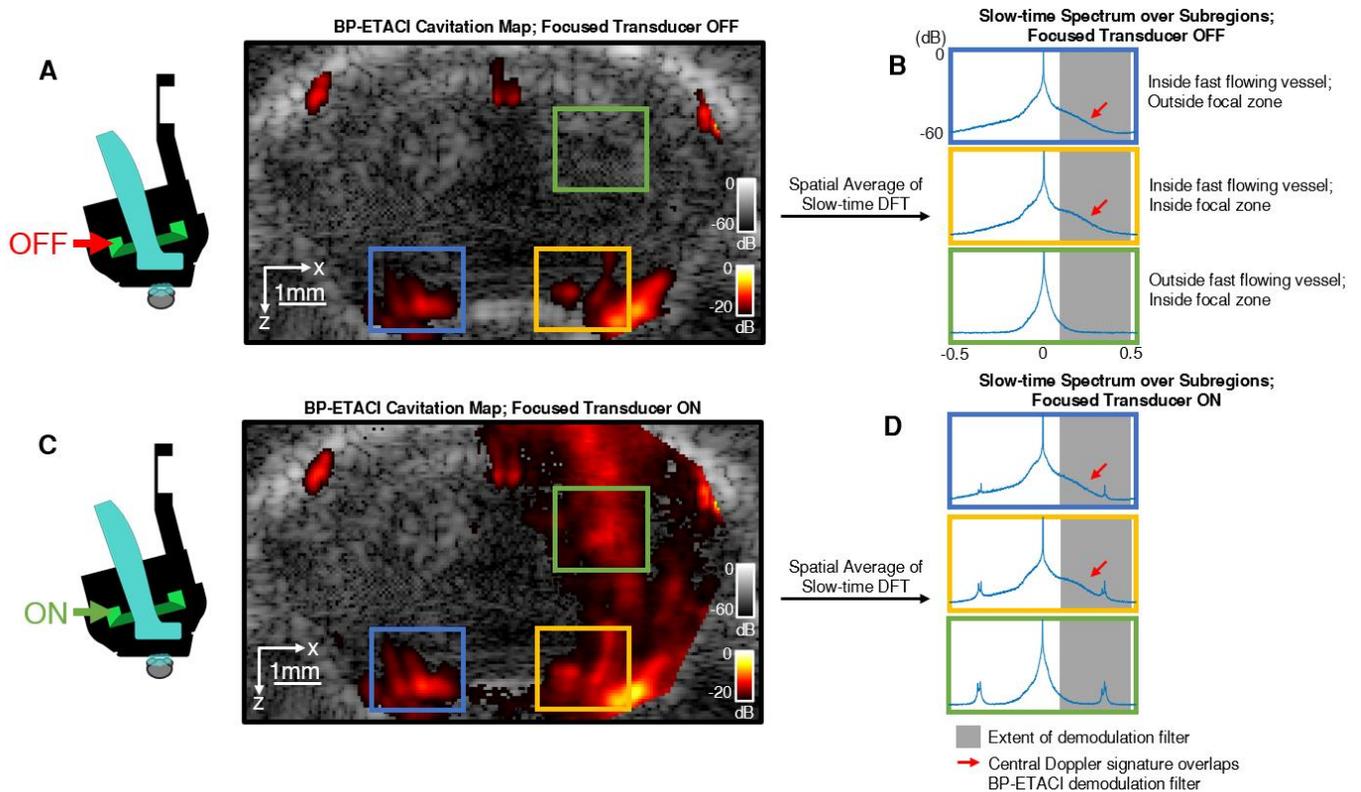

Fig. 7. BP-ETACI cavitation mapping. (A) BP-ETACI cavitation map created when the focused transducer is OFF. Signal is only visible in vessels where the PD spectrum overlaps with the BP-ETACI spectrum. (B) Slow-time spectrum averaged over selected subregions in the field of view. (C) BP-ETACI cavitation map created from data when the focused transducer is ON. The focal zone becomes visible. (D) Slow-time spectrum averaged over selected subregions in the field of view.

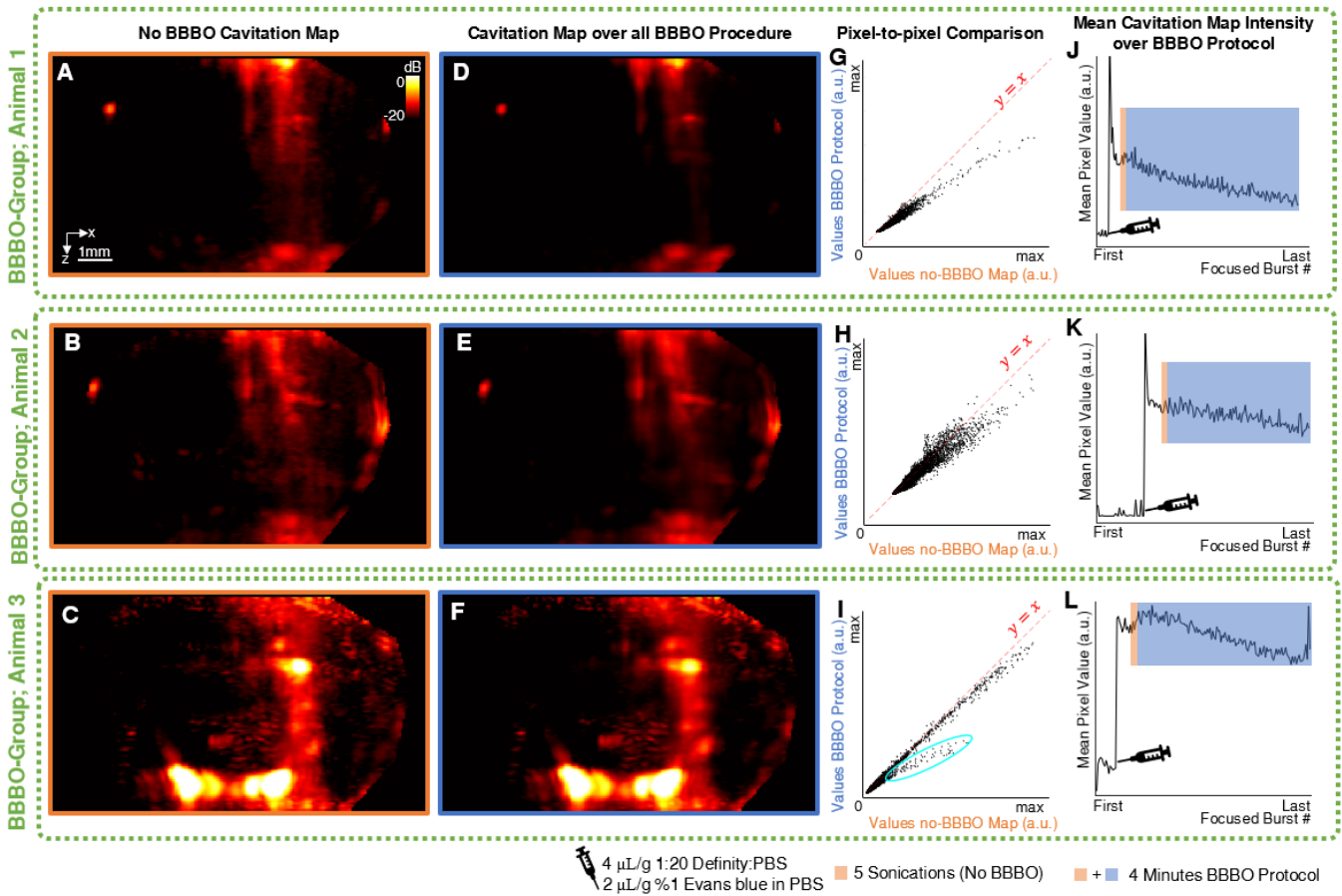

Fig. 8. BP-ETACI-produced cavitation maps show the same structures when built from doses the do not induce BBBO as when built from data acquired over a BBBO procedure. (A-C) Cavitation maps produced with doses that do not open the BBBO. (D-F) Cavitation maps produced over an entire BBBO procedure. (G-I) Pixel to pixel comparison between non-BBBO and BBBO cavitation maps. (J-L) Mean BP-ETACI intensity over the therapeutic focused bursts. The bursts used to build the non-BBBO and the BBBO cavitation maps are highlighted, and the moment µB were injected is indicated.

### *BP-ETACI Cavitation Mapping Predicts BBBO Location In Vivo*

BP-ETACI cavitation mapping showed cavitation regions in the ipsi-lateral side of the brain (Fig. 9. B-G). Superimposing ULM vascular maps on the BP-ETACI cavitation maps helped identify which parts of the underlying vascular anatomy generated BP-ETACI signal. There was strong cavitation in cortical vessels and in large subcortical vessels along the path of the focal zone (Fig. 9. H-M).

Evans blue histology of coronal sections that matched the imaging and therapeutic plane of the animals in the BBBO-group revealed one-sided BBB openings (Fig. 9. N-P). The BBBO region was always situated on the ipsi-lateral side. Overall, the fluorescence was most intense in the top half of the brain, in the cortex and in some subcortical regions. Fluorescence was more intense around vessels, especially in the cortex and subcortical regions. Histology of the animals in the mapping-only-group showed no fluorescence (Fig. 9. Q-S).

The BP-ETACI cavitation maps matched the most intense regions of fluorescence on the histological sections in the cortical and subcortical regions. Low-vascular regions like the corpus callosum and the ventricular system did not generate signal in the cavitation maps and did not show fluorescence in histology. Regions with BP-ETACI signal on the contra-lateral side of the cavitation maps were void of fluorescence in histology and thus contra-lateral BP-ETACI signal and BBBO were uncorrelated.

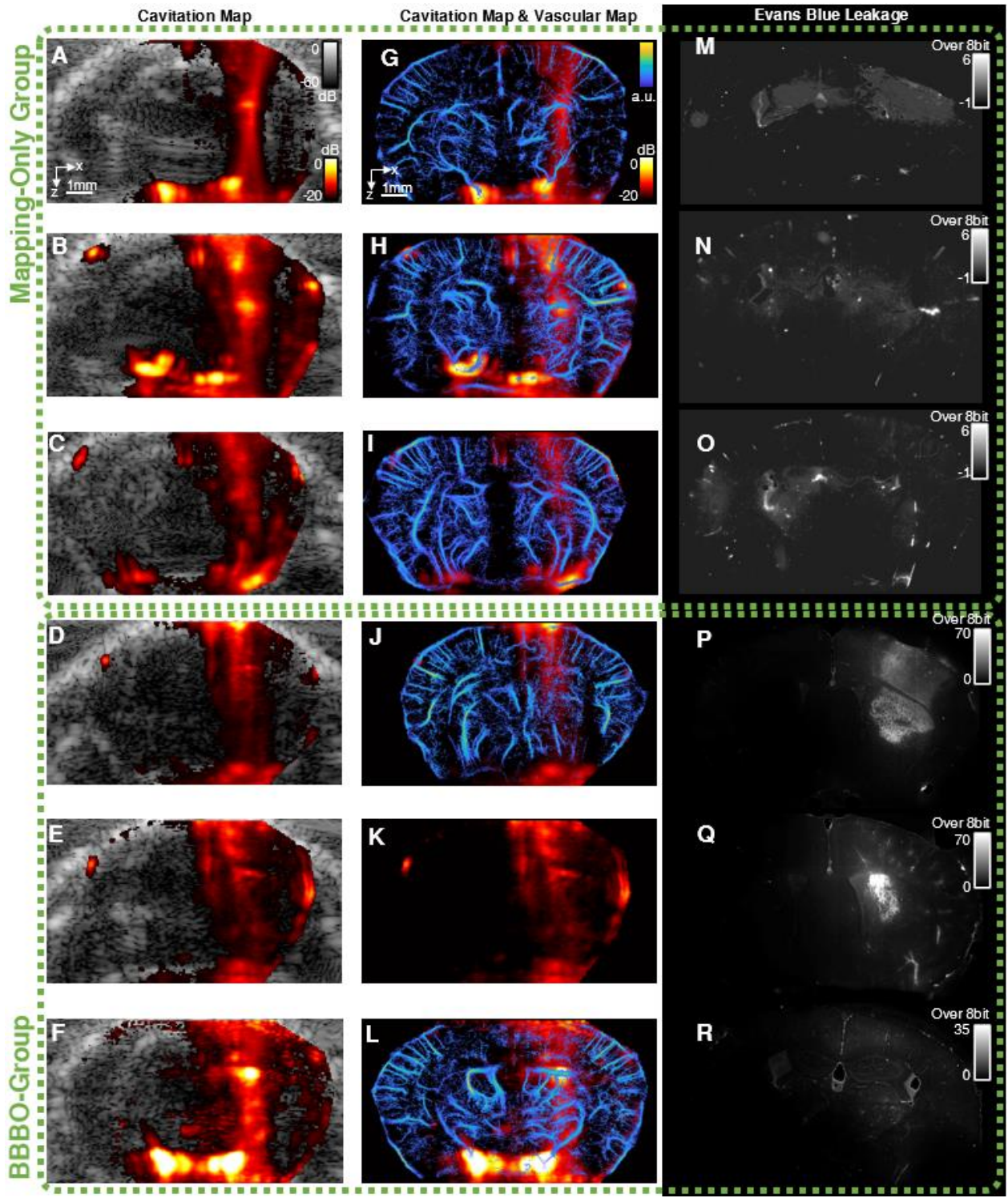

Fig. 9. Transcranial BBBO prediction in the in vivo mouse with BP-ETACI cavitation maps. (A-F) Cavitation maps (red) produced in each animal of the study overlayed on a B-mode image of the skull (gray). (G-L) Correspondence between the cavitation map (red) and the ULM-mapped vasculature (Blue). The ULM vascular map for K was not available. (M-R) Evans blue fluorescence showing the extent of the delivery area. The dynamic range was adjusted for each fluorescence image to show either single sided openings or the absence of opening. (M-O) In the mapping-only group, no Evans blue extravasation. (P-R) in the BBBO-group, single sided Evans blue extravasation consistent with the position of the focal zone.

# DISCUSSION

In this study, we evaluated the potential of BP-ETACI to map the non-inertial cavitation in the mouse brain in vivo to support FUS-BBBO planning. A sampling framework based on bandpass sampling to separate the cavitating µB from the rest of the signal was developed and validated in vitro. Precise aliasing allowed BP-ETACI to place the MHz range cavitation of µB away from unmodulated PD in the frequency domain by using a physically attainable framerate in ultrafast pulse-echo imaging. In vitro, tissue and non-cavitating µB did not produce BP-ETACI signal, while cavitating µB produced strong BP-ETACI signal at the expected location in the spectrum. In the mouse brain, BP-ETACI predicted BBBO location while requiring cavitation doses lower than the BBBO threshold.

The bandpass signal framework developed in this study allows for either the cavitation frequency or the frame rate to be adjusted to offset the cavitation signal from the unmodulated PD signal in the frequency domain. Previous studies have used both approaches under a paradigm that required the PRP and the cavitation period to have a common multiple which enabled ultrafast planewave acquisition as the steering angle could be changed between RM cycles. Muleki-Seya et al. [20] pioneered this paradigm and manipulated the cavitation frequency to create a beat with a set PRP. Instead of the expected single beat frequency, multiple frequencies were observed. All generated frequency bands were demodulated and summed to reconstruct contrast images. Jing and Lindsey [21] used a set, very low cavitation frequency of 100 kHz and were able to induce RM at depth through ex vivo bone. The 1 µs PRF quantum of their programmable ultrasound system was sufficient to sample the RM cycle at 10 different phases by adding a wait period of 1 us to the PRP between frames. In a previous study with ETACI, Blais et al. [15] used a set RM frequency and manipulated the framerate by increasing the transmit beamforming delays frame-to-frame, which allowed the use of the higher frequency cavitation required for BBBO in the mouse brain. The ultrafast paradigm used in these studies required the use of commonly multipliable PRP and cavitation periods. We believe that the reliance on a whole number of samples per RM cycle, and the precision in the timings required to do so, was why studies such as [20], [21] and [15] found frequency peaks not predicted by theory. A critical difference when reframing the sampling process as bandpass sampling is that it allowed the placement of the BP-ETACI aliases anywhere in the spectrum and it eliminated the constraining relationship between PRF and modulation frequency. It yielded a spectrum with a single modulated band on each side of the central PD band and made in vivo use possible as it allowed more frequency-space to separate unmodulated PD and cavitation.

BP-ETACI, like PD, is a pulse-echo technique where the signal is sampled in slow time for each pixel of the image. PD techniques enabled high resolution blood flow imaging by decoupling the imaging resolution from the blood-flow signal sampling. BP-ETACI inherits this advantage from PD imaging but applies it to cavitation imaging instead. As such, BP-ETACI's axial resolution is dictated by the imaging pulse and sequence, instead of being limited by the reception aperture like in PAM. The technique was also applied in real-time, which was used to monitor the targeting of the focal zone. We showed that BP-ETACI resolves the structure of brain tissue and the regional vascular topology that contains the cavitating µB. Resolving such structures is important because it has been demonstrated that vessels seed molecular extravasation in FUS-BBBO [25]. Indeed, in the current study, Evans blue fluorescence was most intense around vessels and was less intense in the corpus callosum.

Prior work has found that the size and density of vessels especially matter in FUS-BBBO [10]. Cavitation mapping at a vascular scale meant that BP-ETACI and Evans blue fluorescence represent different aspects of FUS-BBBO. BP-ETACI mapped the cavitation while fluorescence showed Evans blue extravasation and cellular uptake in the targeted region. BP-ETACI could be used to further study therapeutic agent kinetics after BBBO. Indeed, high-resolution cavitation imaging techniques like BP-ETACI make it possible to predict the regions reached by the delivered molecule and help planning for treatment to reduce off-target effects and increase therapeutic dose in FUS-BBBO [26].

The images presented in the current study were distorted because the experimental assembly used a layer of water at least 100 wavelengths thick above the target, which especially affected the ULM images. Because the relevant vasculature was still observable, no attempt was made to correct the difference in sound speed between the water and the target. However, ULM images and BP-ETACI cavitation maps could benefit from aberration correction techniques [27], particularly when adapted to ultrafast imaging [28].

The BP-ETACI cavitation maps contained unmodulated PD signal from vessels undergoing fast flow. In principle, these areas could be detected beforehand and subtracted from the BP-ETACI maps. BP-ETACI still provided useful information as fast vessels did not overlap with the focal zone. Furthermore, they were easily identifiable with knowledge about the anatomy of the large, symmetrical, high-flow arteries of the brain. The current study made no attempt to eliminate this flow-generated signal, but it could be eliminated by increasing the framerate and using a scheme where the modulated bands are aliased to $f/f_s = 0.5$ to maximize the frequency separation between the modulated bands and the central PD band.

Long (300ms) tone bursts were used as the FUS sonication to create the cavitation maps in this study, allowing acquisition of ensembles of 1200 frames during a single focused transducer burst, which was optimal for the data transfers of the programmable ultrasound system. However, BP-ETACI does not require long bursts. Short therapeutic bursts of 10-50 ms, which are consistent with existing literature on FUS-BBBO [29] are most attainable as long as enough frames are acquired during a burst to allow the detection of the cavitation. However, short, few-cycles therapeutic pulses present a challenge. Further optimization on the framerate, the ensemble size and the demodulation filter transient time are possible.

## CONCLUSION

BP-ETACI was used as a transcranial non-inertial cavitation mapping method for FUS-BBBO in the mouse brain. Using a novel RM framework based on bandpass sampling, we induced separation between motion and cavitation frequency bands. Cavitation mapping using this framework revealed cavitation-bearing vasculature at high resolution that matched vascular maps. Cavitation mapping was possible at non-BBBO doses and predicted the BBBO location, with ipsi-lateral cavitation signals correlating with Evans blue extravasation. The developed technique is easy to use and is directly compatible with broad sonication parameters in FUS-BBBO experiments. The high-resolution cavitation maps can be obtained in real time transcranially in mice, paving the way for further studying FUS-BBBO in the treatment of neurological diseases.

## ETHICS

Animal handling was done in observation of the Canadian Council on Animal Care guidelines and in accordance with McGill University Animal Care Committee regulation under protocol 4532.

## FUNDING

We acknowledge the support of FRQNT, TransMedTech, IVADO, CIHR, NSERC (DGECR-2020-00229) and of the CFI (38095 and 246916)